\documentclass[runningheads]{llncs}
\input{stefmac.sty}
\usepackage{graphics,graphicx,color,latexsym,amssymb,amsfonts,verbatim}

\begin{document}
\pagestyle{headings}\setcounter{page}{1}

\mainmatter

\title{A sound spatio-temporal Hoare logic for the verification of structured interactive programs with
  registers and voices\vspace{-.3cm}\thanks{
This research was partially supported by the Romanian Ministry of Education and Research (PNCDI-II Program 4,
Project 11052/18.09.2007: {\em GlobalComp - Models, semantics, logics and technologies for global
  computing}).}}
\titlerunning{\sc A Hoare Logics for Structured RV-Programs}
\author{Cezara Dragoi\thanks{
Current address: LIAFA, Universite Paris Diderot - Paris 7, France} \and 
Gheorghe Stefanescu\thanks{Current address: Department of Computer Science, 
  University of Illinois at Urbana-Champaign, USA}}
\authorrunning{\sc Dragoi and Stefanescu}
\institute{Faculty of Mathematics and Computer Science, University of Bucharest\\ 
Str. Academiei 14, Bucharest, Romania 010014\vspace{1mm}\\ 
Email: {\tt\normalsize\{cdragoi,gheorghe\}@funinf.cs.unibuc.ro}}
\maketitle\thispagestyle{empty}


\begin{abstract}

Interactive systems with registers and voices (shortly, {\em rv-systems}) are a model for interactive
computing obtained closing register machines with respect to a space-time duality transformation (``voices''
are the time-dual counterparts of ``registers''). In the same vain, AGAPIA v0.1, a structured programming
language for rv-systems, is the space-time dual closure of classical while programs (over a specific type of
data). Typical AGAPIA programs describe open processes located at various sites and having their temporal
windows of adequate reaction to the environment. The language naturally supports process migration, structured
interaction, and deployment of components on heterogeneous machines.\svsp

In this paper a sound Hoare-like spatio-temporal logic for the verification of AGAPIA v0.1 programs is
introduced. As a case study, a formal verification proof of a popular distributed termination detection
protocol is presented.\svsp
\out{
}
\end{abstract}

\section{Introduction}\label{s1}

Verification, a pillar of the development of reliable software, is notoriously difficult. Full verification
was a never reach goal even for sequential programs. (However, currently Floyd-Hoare verification style
regains popularity, being part of modern programming development platforms where users interactively develop
and verify complex software systems.) For concurrent, parallel, or distributed programs, where the tasks are
much more complex, the approach was partially replaced by lighter verification methods (for instance
model-checking, run-time verification, testing, etc.), where specific system properties are verified. The
downside of these propriety-based verification methods is the partial coverage of systems, either due to
theoretical limitations of the methods themselves or due to practical consideration (as the proprietary rights
of certain running platforms). 

Interactive computing \cite{gsv06}\foo{The term ``interactive computation'' often refers to interactive
  systems where one participant is human, dealing with development of powerful human-computer interfaces. In
  our approach, all ``participants'' are ``computing components''.} is a step forward on system
modularization. The approach allows to describe parts of the systems and verify them in an open environment. A
model for interactive computing systems (consisting of interactive systems with registers and voices - {\em
  rv-systems}) and a core programming language (for developing {\em rv-programs}) have been proposed in
\cite{ste06a} based on register machines and a space-time duality transformation. Later on, structured
programming techniques for rv-systems and a kernel programming language AGAPIA have been introduced, with a
particular emphasis on developing a structural spatial programming discipline, see \cite{dr-st07b,dr-st08a}.

Structured process interaction greatly simplifies the construction and the analysis of interactive
programs. For instance, method invocation in current OO-programming techniques may produce unstructured
interaction patterns, with free {\tt goto}'s from a process to another and should be avoided. Compared with
other interaction or coordination calculi (e.g., $\pi$-calculus \cite{mil99}, actor models \cite{agha86}, REO
\cite{arb04}, Orc \cite{mi-co07}, etc.), the rv-systems approach paves the way towards a name-free calculus
and facilitates the development of modular reasoning with good expectations for proof scalability to systems
with thousands of processes.  A new and key element of our structured interaction model is the extension of
temporal data types used on interaction interfaces. These new temporal data types (including voices as a
time-dual version of registers) may be implemented on top of streams similar to the implementation of usual
data types on top of Turing tapes.

AGAPIA \cite{dr-st07b,pss07} is a kernel high-level massively parallel programming language for interactive
computation. It can be seen as a coordination language on top of imperative or functional programming
languages as C++, Java, Scheme, etc.  Typical AGAPIA programs describe open processes located at various sites
and having their temporal windows of adequate reaction to the environment. The language naturally supports
process migration, structured interaction, and deployment of components on heterogeneous
machines. Nonetheless, the language has simple denotational and operational semantics based on scenarios
(scenarios are two-dimensional running patterns; they can be seen as the closure with respect to the
space-time duality transformation of the running paths used to define operational semantics of sequential
programs).

The backbone of our approach to interactive systems is the emphasized space-time duality principle: not only
the model of (structured) rv-systems, but also most of its features or extensions developed so far are all
space-time invariant. For the verification tasks, this duality is again our guiding light towards the
development of Hoare-like spatio-temporal logics for structured rv-programs. We present a rich set of sound
rules $STHlog_0$ for verifying structured rv-programs (no claim on their completeness is included). As a case
study, we present an implementation and a detailed formal verification in $STHlog_0$ of a popular distributed
termination detection protocol. The method may be applied to many other sophisticated distributed protocols. A
short description of the method follows.

For verification of sequential programs we have to find assertions in a few key points of the tested program
and to prove certain invariance conditions, see, e.g., \cite{lics-book}. For rv-programs cut-points become
contours, surrounding finite scenarios. The verification procedure \cite{ste06b} consists of the following
three steps: (i) find an appropriate set of contours and assertions; (ii) fill in the contours with all
possible scenarios; and (iii) prove these scenarios respect the border assertions.  Except for the guess of
assertions, the proof is finite and can be fully automatized. 

The verification of structured rv-programs follows the same pattern. However, structured rv-programs have a
more restricted way to construct scenarios, hence the procedure is more regular: (1) provide assertions for
each basic statement and (2) lift assertions to larger and larger programs applying $STHlog_0$ inference
rules.

The paper is organized as follows. We start with a brief presentation of scenarios and spatio-temporal
specifications. Next, structured rv-programs and a scenario-based operational semantics are presented. A short
section describes the syntax of AGAPIA v0.1 language. Then, our approach for developing Hoare-like
spatio-temporal verification logics is presented. Finally, a detailed proof of the correctness of a
termination detection protocol is included. A brief section on related works conclude the paper.

\section{Scenarios}\label{s-scen}

In this section temporal data, spatio-temporal specifications, grids, scenarios, and operations on scenarios
are briefly presented.

\paragraph{Spatio-temporal specifications.}

What we call ``spatial data'' are just the usual data occurring in imperative programming. For them, common
data structures and the usual memory representation may be used. On the other hand, ``temporal data'' is a
name we use for a new kind of (high-level) temporal data implemented on streams. A {\em stream} \cite{br-st01}
is a sequence of data ordered in time, denoted as $a_0{}^{\frown}a_1{}^{\frown}\dots$ where $a_0,a_1,\dots$
are its elements at time clocks $0,1,\dots$, respectively.  Typically, a stream results by observing data
transmitted along a channel: it exhibits a datum (corresponding to the channel type) at each clock cycle.

A {\em voice} is defined as the time-dual of a register: {\em It is is a temporal data structure that holds a
  natural number. It can be used (``heard'') at various locations. At each location it displays a particular
  value.}

This formulation may be difficult to understand at a first sight (the reader is invited to come back here
after the reading of the section on rv-programs and their scenario semantics). In a different formulation,
this means high-level temporal data structures on streams (including voices) may be common to multiple
processes, each process having particular values for these data structures.

Voices may be implemented on top of a stream in a similar way registers are implemented on top of a Turing
tape, for instance specifying their starting time and their length. Most of usual data structures have natural
temporal representations. Examples include timed booleans, timed integers, timed arrays of timed integers,
etc.

A {\em spatio-temporal specification} $S:(m,p)\ra (n,q)$ (using registers and voices only) is a relation
$S\subseteq (\mathbb{N}^m \times \mathbb{N}^p) \times(\mathbb{N}^n \times\mathbb{N}^q)$, where $m$ (resp. $p$)
is the number of input voices (resp. registers) and $n$ (resp. $q$) is the number of output voices
(resp. registers).  The associated relation $S$ is often functional, sometimes written as $\stdata{v}{r}
\mapsto \stdata{v'}{r'}$, where $v,v'$ (resp. $r,r'$) are tuples of voices (resp. registers).

Specifications may be composed horizontally and vertically, as long as their types agree; e.g., for two
specifications $S_1:(m_1,p_1)\ra (n_1,q_1)$ and $S_2:(m_2,p_2)\ra (n_2,q_2)$ the {\em horizontal composition}
$S_1\hcomp S_2$ is defined only if $n_1=m_2$ and the type of $S_1\hcomp S_2$ is $(m_1,p_1+p_2)\ra
(n_2,q_1+q_2)$; the result is as expected: it consists in tuples $((v,(r_1,r_2)),(v'',(r'_1,r'_2)))$ such that
there exists $v'$ with $((v,r_1),(v',r'_1,))\in f_1$ and $((v',r_2),(v'',r'_2))\in f_2$.

\paragraph{Grids and scenarios.}

\begin{figure}
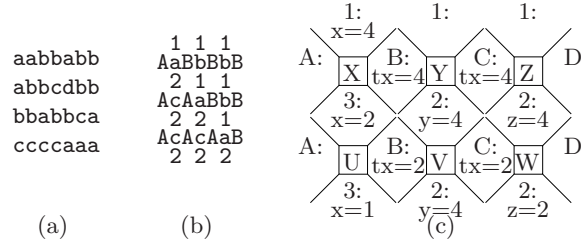
\begin{center}\hspace*{-.5cm}\begin{tabular}{c@{\hsp\hsp}c@{\hsp\hsp\hsp}c}
\raisebox{.7cm}{\tdwbis{small}{aabbabb\\abbcdbb\\bbabbca\\ccccaaa}}
& \raisebox{.9cm}{$\tdwbis{small}{\ 1\ 1\ 1\ \tdret AaBbBbB\tdret \ 2\ 1\
1\ \tdret AcAaBbB\tdret \ 2\ 2\ 1\ \tdret AcAcAaB\tdret 2\ 2\ 2\ }$}
&{\small$\begin{array}{@{}c@{}c@{}c@{}}
\scelli{1:\snvsp\\x=4}{A:}{X}{}{}
&\scelli{1:}{B:\snvsp\\tx=4}{Y}{}{}
&\scelli{1:}{C:\snvsp\\tx=4}{Z}{D}{}
\vspace{-1mm}\\
\scelli{3:\snvsp\\x=2}{A:}{U}{}{3:\snvsp\\x=1}
&\scelli{2:\snvsp\\y=4}{B:\snvsp\\tx=2}{V}{}{2:\snvsp\\y=4}
&\scelli{2:\snvsp\\z=4}{C:\snvsp\\tx=2}{W}{D}{2:\snvsp\\z=2}
\end{array}$}\\
(a)&(b)&(c)\end{tabular}\vspace{-.5cm}\end{center}
\caption{A grid (a), an abstract scenario (b), and a concrete scenario (c).\vspace{-.3cm}}\label{f-gscen}
\end{figure}

A {\em grid} is a {\em rectangular} two-dimensional array containing letters in a given alphabet.  A grid
example is presented in Fig.~\ref{f-gscen}(a). Our default interpretation is that columns correspond to
processes, the top-to-bottom order describing their progress in time. The left-to-right order corresponds to
process interaction in a {\em nonblocking message passing discipline}: a process sends a message to the right,
then it resumes its execution.

A {\em scenario}\foo{See \cite{ha-ma03} for less shape-constrained scenarios.} is a grid enriched with data
around each letter. The data may be given in an abstract form as in Fig.~\ref{f-gscen}(b), or in a more
detailed form as in Fig.~\ref{f-gscen}(c).

The type of a scenario interface is represented as $t_1;t_2;\dots;t_k$, where each $t_k$ is a tuple of simple
types used at the borders of scenario cells.\foo{If only registers and voices are used, then each tuple may by
  simply replace by a number (counting of its components).} The empty tuple is also written 0 or $nil$ and can
be freely inserted to or omitted form such descriptions. The type of a scenario is specified as
$f:\tsrv{w}{n}{e}{s}$, where $w,n,e,s$ are the types for its west, north, east, south interfaces.  For
example, the type of the scenario in Fig.~\ref{f-gscen}(c) is $\tsrv{nil;nil}{sn;nil;nil}
{nil;nil}{sn;sn;sn}$, where $sn$ denotes a spatial integer type.

\paragraph{Operations with scenarios.}

We say two scenario interfaces $t=t_1;t_2;\dots;t_k$ and $t'=t'_1;t'_2;\dots;t'_{k'}$ are {\em equal}, written
$t=t'$, if $k=k'$ and the types and the values of each pair $t_i,t'_i$ are equal. Two interfaces are {\em
  equal up to the insertion of $nil$ elements}, written $t=_nt'$, if they become equal by appropriate
insertions of $nil$ elements.

\begin{figure}\begin{center}
\includegraphics[scale=.40]{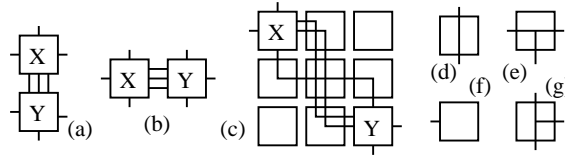}\vspace{-.5cm}\end{center}
\caption{Operations on scenarios}\label{f654}\end{figure}

Let $Id_{m,p}:\tsrv{m}{p}{m}{p}$ denote the constant cells whose temporal and spatial outputs are the same
with their temporal and spatial inputs, respectively; an example is the center cell in
Fig.~\ref{f654}(c), namely $Id_{1,2}$.

\begin{figure}\begin{center}\includegraphics[scale=.33]{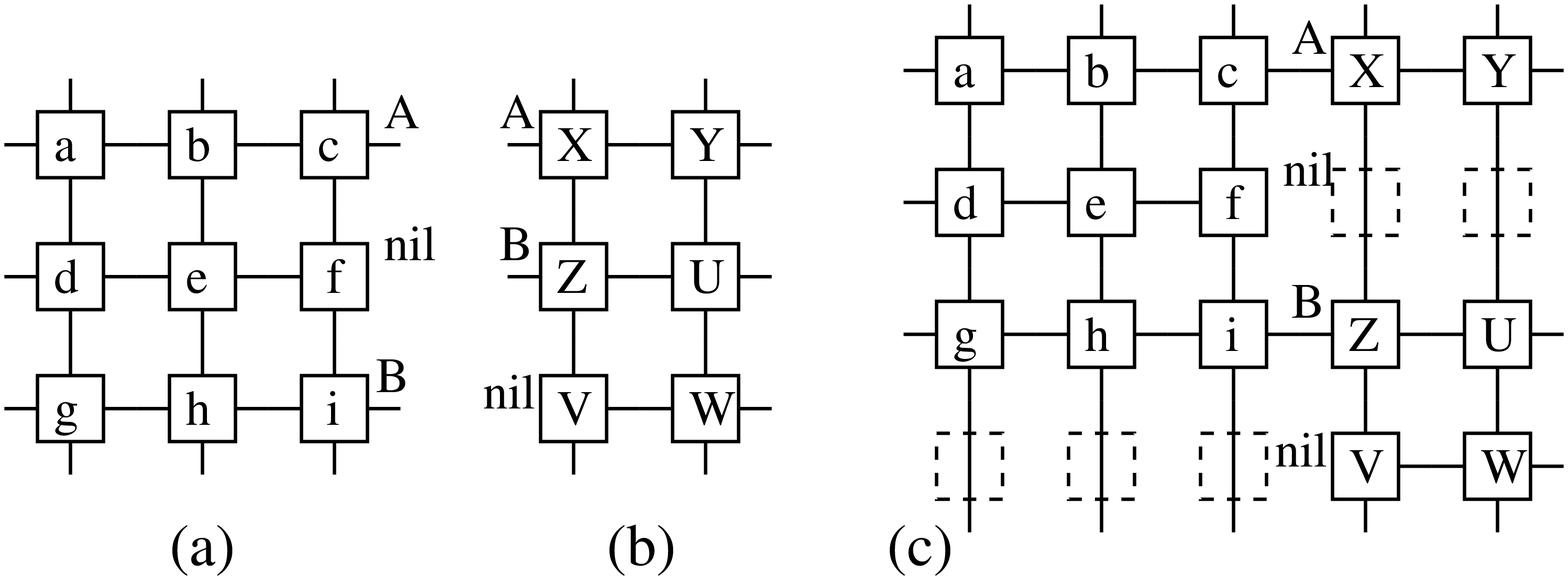}\vspace{-.5cm}\end{center}
\caption{Details for the horizontal composition of scenarios}\label{scenHcomp}\end{figure}

{\em Horizontal composition:} Let $f_i:\tsrv{w_i}{n_i}{e_i}{s_i}, i=1,2$ be two scenarios. Their {\em
  horizontal composition $f_1\hcomp f_2$} is defined only if $e_1=_nw_2$. For each inserted $nil$ element in
an interface (to make the interfaces $e_1$ and $w_2$ equal), a dummy row is inserted in the corresponding
scenario, resulting a scenario $\ol{f_i}$. The result $f_1\hcomp f_2$ is obtained putting $\ol{f_1}$ on left
of $\ol{f_2}$. The operation is briefly illustrated Fig.~\ref{f654}(b) and in more details in
Fig.~\ref{scenHcomp}.  The result is unique up to insertion or deletion of dummy rows. Its identities are
$Id_{m,0}, m\geq 0$.

{\em Vertical composition:} The definition of {\em vertical composition $f_1\vcomp f_2$} (see
Fig.~\ref{f654}(a)) is similar, but now $s_1=_nn_2$. For each inserted $nil$ element (to make $s_1$ equal to
$n_2$), a dummy column is inserted in the corresponding scenario, resulting a scenario $\ol{f_i}$. The result
$f_1\vcomp f_2$ is obtained putting $\ol{f_1}$ on top of $\ol{f_2}$.  Its identities are $Id_{0,m}, m\geq 0$.

{\em Diagonal composition:} This is a derived operation. The {\em diagonal composition $f_1\dcomp f_2$} (see
Fig.~\ref{f654}(c)) is defined only if $e_1=_nw_2$ and $s_1=_nn_2$. The result is defined by the formula\bi
\item[] $f_1\dcomp f_2 =(f_1\hcomp R_1\hcomp \Lambda)\vcomp(S_2\hcomp Id \hcomp R_2)\vcomp(\Lambda \hcomp
  S_1\hcomp f_2).$\ei for appropriate constants $R,S,Id,\Lambda$. Its identities are $Id_{m,n}, m,n\geq
  0$. (The involved constants $R,S,Id,\Lambda$ are described below.)

{\em Constants:} Except for the defined identities, we use a few more constants. Most of them may be found in
Fig.~\ref{f654}(c): a recorder $R$ (2nd cell in the 1st row), a speaker $S$ (1st cell in the 2nd row), an
empty cell $\Lambda$ (3rd cell in the 1st row). Other constants of interest are: transformed recorders
Fig.~\ref{f654}(e) and transformed speakers Fig.~\ref{f654}(g).

\section{Structured rv-programs}\label{s-srv}

Rv-programs \cite{ste06a} resemble flowcharts and assembly languages: one can freely uses {\tt go-to}
statements with both, temporal labels (free jumps in a process from a statement to another) and spatial labels
(free jumps in a macro-step from a process to another). The original approach of the authors in 2006 was to
introduce structured programming techniques on top of rv-programs.  However, the resulted structured
rv-programs and their scenario based semantics may be described directly, from scratch. The lower level of
rv-programs is still useful (and important!), as it may be used as a target language for compiling and is more
appropriate for running programs on (possible multicore/manycore) computer architectures. Below, we restrict
ourself to structured rv-programs. The Hoare logics for structured rv-programs, to be presented later in the
paper, has its roots in a Floyd logics developed for unstructured rv-programs \cite{ste06b}.

\paragraph{The syntax of structured rv-programs.}

The basic blocks for constructing structured rv-programs are {\em modules}. A module gets input data from its
west and north interfaces, process them (applying the module's code), and delivers the computed outputs at its
east and south interfaces. While one can argue for the use of nondeterministic behaviors for processes
associated to modules, we afraid of doing so: in all the example we have developed so far the basis modules
(and the resulting structured rv-programs) have deterministic behavior. Nondeterministic behaviors naturally
occur at more abstract levels when lot of details on particular low-level temporal or spatial data are hidden.

On top of modules, structured rv-programs are built up using ``if'' and both, composition and iterated
composition statements for the vertical, the horizontal, and the diagonal directions. The composition
statements capture at the program level the corresponding operations on scenarios. The iteration statements
are also called the {\em temporal}, the {\em spatial}, and the {\em spatio-temporal while statements} - their
scenario meaning is described below.

The {\em syntax for structured rv-programs} is given by the following BNF grammar\bi\item[] $P::=
{}$\parbox[t]{8cm}{$X\ |\ if(C)then\{P\}else\{P\} |\ P\pvcomp P\ |\ P\phcomp P\ |\ P\pdcomp
  P\\ |\ while\_t(C)\{P\}\ |\ while\_s(C)\{P\} |\ while\_st(C)\{P\}$}\svsp\\ $X::=
{}$\parbox[t]{8cm}{$module\{listen\ t\_vars\}\{read\ s\_vars\}\\\{code;\}\{speak\ t\_vars\}\{write\ s\_vars\}$}\ei
This is a core definition of structured rv-programs, as no data types or language for module's code is
specified. On the other hand, Agapia, to be shortly presented, is a concrete incarnation of structured
rv-programs into a fully running environment.

Notice that we use a different notation for the composition operators on scenarios $\vcomp,\hcomp,\dcomp$ and
on programs $\pvcomp,\phcomp,\pdcomp$. Moreover, to avoid confusion, the extension of the usual program
composition operator ';' to structured rv-programs (i.e., the vertical composition) is denoted by a different
symbol ``$\pvcomp$''.

\paragraph{Operational semantics.}

The operational semantics \snvsp$$|\ \ |:\mbox{Structured rv-programs}\ra \mbox{Scenarios}\snvsp$$ associates
to each program the set of its running scenarios.

The type of a program $P$ is denoted $P:\tsrv{w(P)}{n(P)}{e(P)}{s(P)}$, where $w(P)/n(P)/e(P)/s(P)$ indicate
its types at the west/north/east/south borders. On each border, the type may be quite complex (see AGAPIA
interface types in Sec.~\ref{s-agap}). The convention is to separate by ``,'' the data from within a module
and by ``;'' the data coming from different modules. This convention refers to both spatial and temporal data.

The type associated to a program may include different types for the interfaces of its running scenarios. For
instance, a temporal while statement may have running scenarios with different numbers of rows which may
exhibit different interfaces at their west/east borders. With this explanation, the definition below makes
sense. We say, two interface types {\em match} if they have a nonempty intersection.

\paragraph{Modules.} 

Modules are the starting blocks for building structured rv-programs. The {\tt listen (read)} instruction is
used to get the temporal (spatial) input and the {\tt speak (write)} instruction to return the temporal
(spatial) output. The {\tt code} consists in simple instructions as in the C code. No distinction between
temporal and spatial variables is made within a module.

A scenario for a module consists of a unique cell, with concrete data on the borders, and such that the output
data are obtained from the input data applying the module's code.

\paragraph{Composition.} 

\begin{figure}\begin{center}\begin{tabular}{c@{\hspace{1cm}}c@{\hspace{1cm}}c}
\includegraphics[scale=.4]{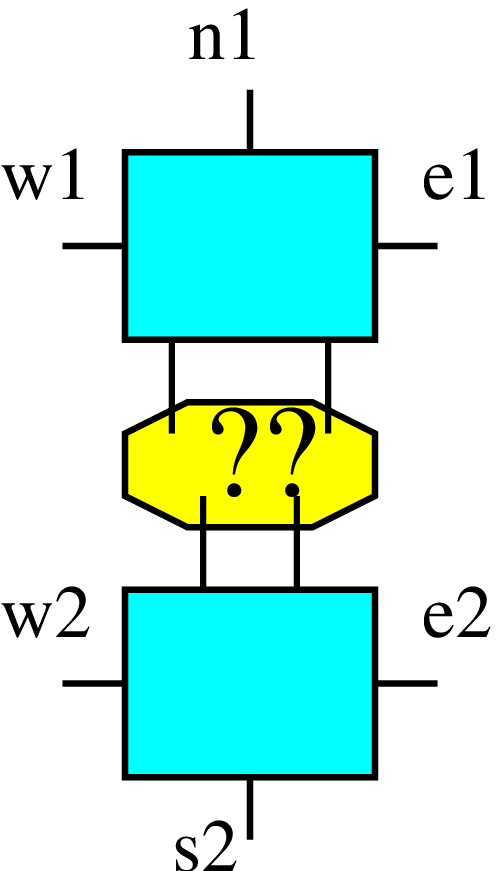} & \raisebox{.75cm}{\includegraphics[scale=.4]{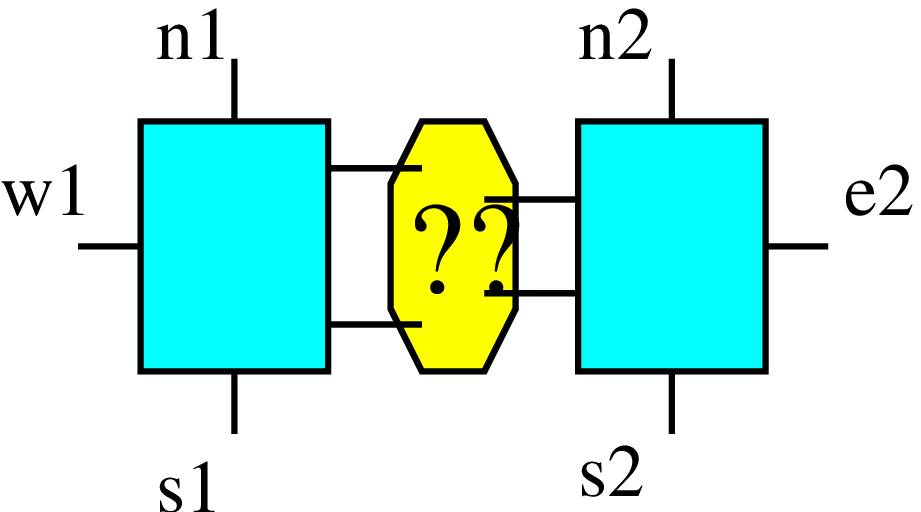}}
& \raisebox{.4cm}{\includegraphics[scale=.4]{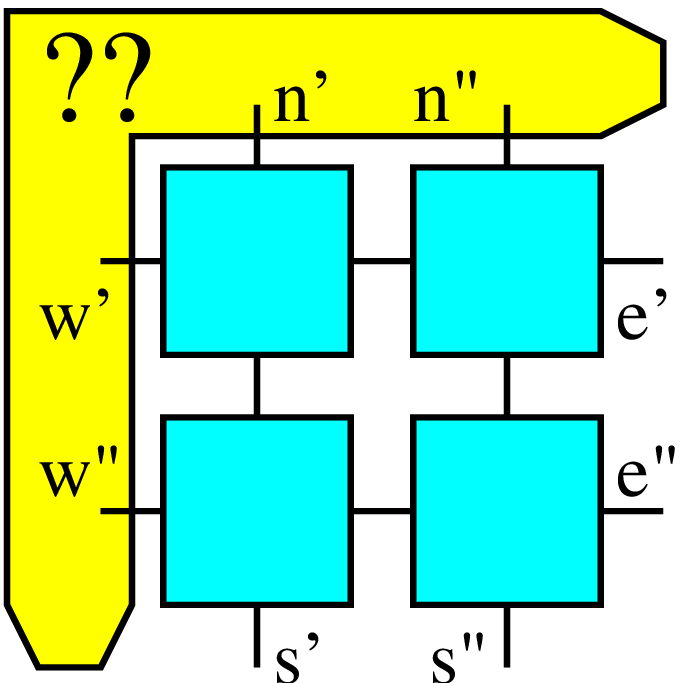}}\vspace{-.75cm}
\end{tabular}\end{center}\caption{Vertical/horizontal compositions and ``if'' statement}\label{vh+if}\end{figure}

Programs may be composed ``horizontally'' and ``vertically'' as long as their types on the connecting
interfaces agree. They can also be composed ``diagonally'' by mixing the horizontal and vertical compositions.

For two programs $P_i:\tsrv{w_i}{n_i}{e_i}{s_i}$, $i=1,2$ we define the following composition operators.

{\em Horizontal composition:} $P_1\phcomp P_2$ is defined if the interfaces $e_1$ and $w_2$ match, see
Fig.~\ref{vh+if}(left). The type of the composite is $\tsrv{w_1}{n_1;n_2}{e_2}{s_1;s_2}$. A scenario for
$P_1\phcomp P_2$ is a horizontal composition of a scenario in $P_1$ and a scenario in $P_2$.

{\em Vertical composition:} $P_1\pvcomp P_2$ is similar.

{\em Diagonal composition:} $P_1\pdcomp P_2$ is defined if $e_1$ matches $w_2$ and $s_1$ matches $n_2$. The
type of the composite is $\tsrv{w_1}{n_1}{e_2}{s_2}$. A scenario for $P_1\pdcomp P_2$ is a diagonal
composition of a scenario in $P_1$ and a scenario in $P_2$.

\paragraph{If.}

For two programs $P_i:\tsrv{w_i}{n_i}{e_i}{s_i}$, $i=1,2$, a new program $Q=if\ (C)\ then\ P_1\ else\ P_2$ is
constructed, where $C$ is a condition involving both, the temporal variables in $w_1\cap w_2$ and the spatial
variables in $n_1\cap n_2$, see Fig.~\ref{vh+if}(right). The type of the result is $Q:\tsrv{w_1\cup
  w_2}{n_1\cup n_2}{e_1\cup e_2}{s_1\cup s_2}$.

A scenario for $Q$ is a scenario of $P_1$ when the data on the west and the north borders of the scenario
satisfy condition $C$, otherwise is a scenario of $P_2$ (with these data on the borders).

\paragraph{While.} 

Three types of while statements are used for defining structured rv-programs, each being the iteration of a
corresponding composition operation.

{\em Temporal while:} For a program $P:\tsrv{w}{n}{e}{s}$, the statement $while\_t\ (C)\{P\}$ is defined if
the interfaces $n$ and $s$ match and $C$ is a condition on the spatial variables in $n\cap s$. The type of the
result is $\tsrv{(w;)^*}{n\cup s}{(e;)^*}{n\cup s}$. A scenario for $while\_t\ (C)\{P\}$ is either an
identity, or a repeated vertical composition $f_1\vcomp f_2\vcomp\dots\vcomp f_k$ of scenarios for $P$ such
that: (1) the north border of each $f_i$ satisfies $C$ and (2) the south border of $f_k$ does not satisfy $C$.

{\em Spatial while:} $while\_s\ (C)\{P\}$ is similar.

{\em Spatio-temporal while:} For $P:\tsrv{w}{n}{e}{s}$, the statement $while\_st\ (C)\{P\}$ is defined if $w$
matches $e$ and $n$ matches $s$ and, moreover, $C$ is a condition on the temporal variables in $w\cap e$ and
the spatial variables in $n\cap s$. The type of the result is $\tsrv{w\cup e}{n\cup s}{w\cup e}{n\cup s}$. A
scenario for $while\_st\ (C)\{P\}$ is either an identity, or a repeated diagonal composition $f_1\dcomp
f_2\dcomp\dots\dcomp f_k$ of scenarios for $P$ such that: (1) the west and north border of each $f_i$
satisfies $C$ and (2) the east and south border of $f_k$ does not satisfy $C$.

A few particular cases of while statement may be easier to understand and use. For instance, when the body
program $P$ of a temporal while statement has dummy temporal interfaces, the temporal while coincides with the
while from imperative programming languages.

\section{The AGAPIA v0.1 programming language}\label{s-agap}

\begin{figure}[t]\parbox[t]{7cm}{\footnotesize
{\bf Interfaces}\svsp\\
$SST::={}$ \parbox[t]{7cm}{$nil\ |\ sn\ |\ sb\ |\ (SST\cup SST)\\ |\ (SST,SST)\ |\
(SST)^*$}\svsp\\
$ST::={}$ \parbox[t]{7cm}{$ (SST)\ |\ (ST\cup ST)\\ |\ (ST;ST)\ |\ (ST;)^*$}\svsp\\
$STT::={}$ \parbox[t]{7cm}{$nil\ |\ tn\ |\ tb\ |\ (STT\cup STT)\\ |\ (STT,STT)\ |\
(STT)^*$}\svsp\\
$TT::= {}$ \parbox[t]{7cm}{$(STT)\ |\ (TT\cup TT)\\ |\ (TT;TT)\ |\ (TT;)^*$}\vsp

{\bf Expressions}\svsp\\
$V::={}$ \parbox[t]{7cm}{$ x:ST\ |\ x:TT\ |\ V(k)\\ |\ V.k\ |\ V.[k]\ |\ V@k\ |\ V@[k]$}\svsp\\
}\hspace*{-1.5cm}\parbox[t]{9cm}{\footnotesize
$E::= n\ |\ V\ |\ E+E\ |\ E*E\ |\ E-E\ |\ E/E$\svsp\\
$B::= b\ |\ V\ |\ B\&\& B\ |\ B||B\ |\ !B\ |\ E<E$\vsp\\
{\bf Programs}\svsp\\
$W::={}$ \parbox[t]{7cm}{$nil\ |\ new\ x:SST\ |\ new\ x:STT\\ |\ x := E\ |\ if (B) \{W\} else \{W\}
\\ |\ W;W\ |\ while (B) \{W\}$}\svsp\\
$M::={}$ \parbox[t]{7cm}{$module\{listen\ x:STT\}\{read\ x:SST\}\\
\{W;\}\{speak\ x:STT\}\{write\ x:SST\}$}\svsp\\
$P::={}$ \parbox[t]{7cm}{$nil\ |\ M\ |\ if (B) \{P\} else \{P\}\ \\ 
|\ P\pvcomp P\ |\ P\phcomp P\ |\ P\pdcomp P\ \\
|\ while\_t (B) \{P\}\ |\ while\_s (B) \{P\}\\ |\ while\_st (B) \{P\}$}}
\caption{The syntax of AGAPIA v0.1 programs}\label{f-agapia}
\end{figure}

To develop and verify structured rv-programs for concrete computation tasks we need at least a couple of basic
data types. The AGAPIA v0.1 programming language \cite{dr-st07b}, to be be shortly introduced, forms a minimal
languages: it describes what is obtained allowing for spatial and temporal integer and boolean types and
applying structured rv-programming statements.

The syntax for AGAPIA v0.1 programs is presented in Fig.~\ref{f-agapia}. The v0.1 version is intentionally
kept simple to illustrate the key features of the approach (see \cite{pss07} for v0.2 extension, including
high-level structured rv-programs). The language is space-time invariant\foo{This means, we can formally
  define a space-time duality operator which maps an AGAPIA v0.1 program $P$ to another AGAPIA v0.1 program
  $P^{\vee}$ such that $P=P^{\vee\vee}$.} and has global scoping within modules and local scoping outside.

The types for spatial interfaces are built up starting with integer and boolean $sn,sb$ types, applying the
rules for $\cup,\mbox{','},(\_)^*$ to get process interfaces, then the rules for $\cup,\mbox{';'},(\_;)^*$ to
get system interfaces. The temporal types are similarly introduced. Given a type $V$, the notations
$V(k),V.k,V.[k], V@k,V@[k]$ are used to access its components.\foo{See \cite{dr-st07b} for details - we will
  not use this notation in the present paper.} Expressions, usual while programs, modules, and programs are
then naturally introduced. Notice that AGAPIA v0.1 has a strongly restricted format: {\em module and program
  statements are not mixed} (in the new v0.2 version of AGAPIA \cite{pss07} programs have no longer this
restriction - modules and programs can be freely combined).

An useful derived statement, to be used in the next sections, is a spatial ``for'' statement {\tt
  for\_s(i=a;i<b;i++)\{R\}}. This is a macro stating for {\tt
  i=a\phcomp\ while\_s(i<b)\{R\phcomp\ i++\phcomp\}}, where {\tt i=a} and {\tt i++} denote modules with such
code, with empty spatial interfaces, and whose temporal interfaces are equal to the temporal interface of {\tt
  R} (where {\tt i} is included).

\section{Towards a Hoare-like logic for structured rv-programs}\label{s-hoare}

This section describes an approach for developing verification logics for structured rv-programs.  The
presentation starts with a few words on the verification of unstructured rv-programs (more details on
developing Floyd logics for unstructured rv-programs may be found in \cite{ste06b}).

The semantics of (structured) rv-programs uses scenarios, a two-dimensional version of running paths used in
sequential programs. The lifting of Floyd verification method to rv-programs is essentially a two-dimensional
extension where cut-points with assertions become contours (borders of certain scenarios) with appropriate
assertions.

\paragraph{The method.}

The Floyd method for unstructured sequential programs, requires to find assertions in a few key points of the
programs and to prove appropriate invariance conditions. It should be at least one cut-point along each loop.
The set of cut-points ensures that: (1) each syntactically possible path from input to output is decomposed
into a sequence of small paths $p_1p_2\dots p_k$, each $p_i$ tarting and ending with cut-points and containing
no cut-point inside and (2) the set of all these $p_i$ forms a {\em finite set} $K$.  The proof finally
reduces to the verification of the invariance conditions for the paths in $K$.

For rv-programs, cut-points becomes contours surrounding finite scenarios. Their set must be finite.  The
condition to ``break all loops'' becomes ``each syntactically possible scenario can be decomposed in pieces
corresponding to these contours''. To conclude, the verification procedure for rv-programs consists of the
following three steps: (i) find an appropriate set of contours and assertions; (ii) fill in the contours with
all possible scenarios; and (iii) prove these scenarios respect the border assertions.  Notice that, except
for the guess of good assertions, the proof is finite and can be fully automatized.

Structured rv-programs have a more restricted way to construct scenarios, hence the procedure is expected to
be more regular: (1) provide assertions for each basic statement and (2) use appropriate inference rules to
lift the assertions to larger programs.

\paragraph{Hoare-assertions.}

As we said, structured rv-programs have a restricted way to form scenarios, hence one expects the assertion
format may be somehow simplified. However, notice that we are working in an open environment, hence the local
application of a rule is to be integrated into a larger context, including assertions on parts of the contour
that may look irrelevant to the current piece of code\foo{An example is P2, the invariant used in the
  verification of the termination detection protocol in the next section.}, but are needed to infer the
correctness of the behavior of the full system.

\begin{figure}\begin{center}\begin{tabular}{cc}
\includegraphics[scale=.4]{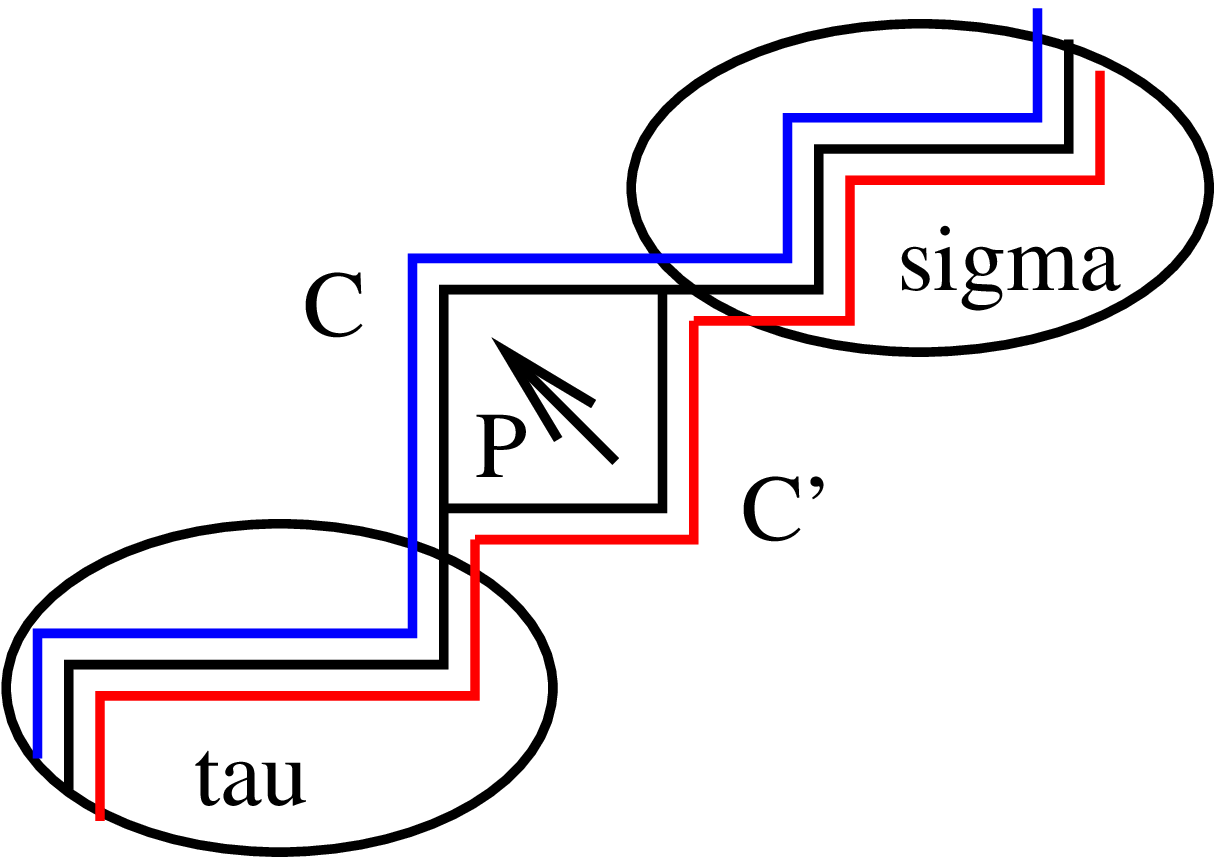} & \hspace*{-.5cm}\includegraphics[scale=.4]{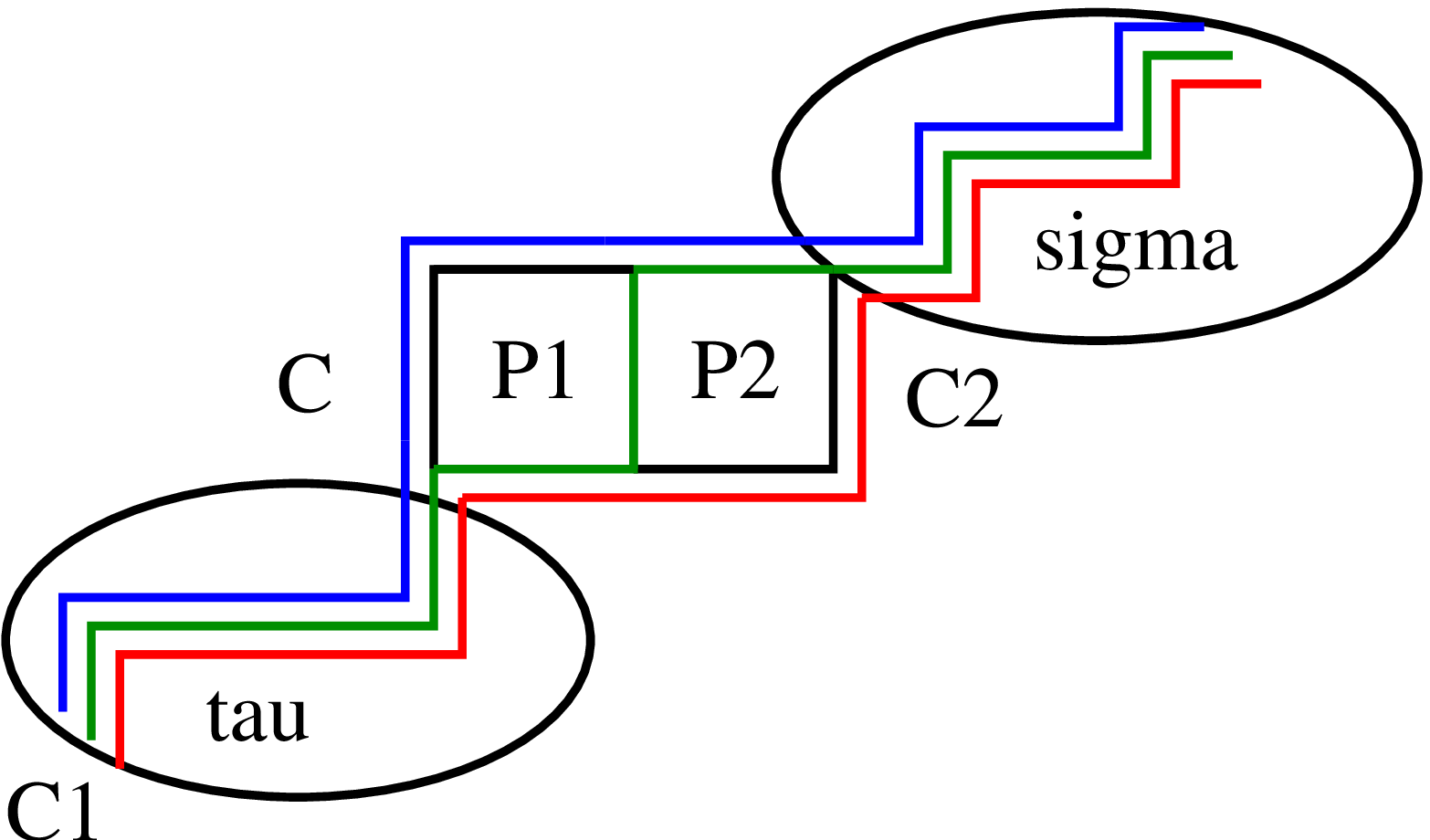}
\end{tabular}\end{center}
\caption{Illustrations for the ``Basic Rule'' and the ``Rule for horizontal composition''}
\label{ro-r1}\end{figure}

An assertion (for structured rv-programs) is defined using a rectangular contour surrounding a piece of
structured rv-program and extended with dummy contours from its top-right and bottom-left corners, loosely
along the 2nd diagonal. (Contours and assertions, to be shortly defined, are illustrated in Fig.~\ref{ro-r1}.)
Formally, a {\em Hoare-assertion contour} is defined by a pair of lines starting from the same point
\snvsp$$\tau N^kE^l\sigma, \tau E^lN^k\sigma\snvsp$$ where $N$ and $E$ denote unit lines towards the north and
the east directions, and $\tau$ and $\sigma$ are sequences of lines of the following types:
$N^{a_1}E^{b_1}\dots N^{a_k}E^{b_k}$, $E^{b_1}\dots N^{a_k}E^{b_k}$, $N^{a_1}E^{b_1}\dots N^{a_k}$,
$E^{b_1}\dots N^{a_k}$, where all $a_i,b_j\geq 1$. Assertions use variables on the contour border. A border
unit line is either horizontal and has an index from left, or vertical and has an index from top.\foo{Notice
  that the index is not changed when a program is applied. However, it may be changed when one insert or
  delete lines with $nil$ type on the appropriate borders to handle program compositions.}  The variables for
the unit border lines are refereed to using these indices.

A {\em Hoare assertion} (see Fig.~\ref{ro-r1}(left)) is a formula \snvsp\bi\item[]\hspace*{1cm}\hass{\tau
  (N^kE^l)\sigma\mbox{:} C}{P}{\tau (E^lN^k)\sigma\mbox{:} C'}\snvsp\ei where $C$ is a condition on the
west-north part $\tau (N^kE^l)\sigma$ of the contour, $P$ is a structured rv-program, and $C'$ is a condition
on the south-east part $\tau (E^lN^k)\sigma$ of the contour. The conditions $C,C'$ are first-order formulas
described using contour variables. The pair of parentheses $(..)$ locates the part of the contour where
program $P$ is used.

\paragraph{Inference rules.}

We consider the following set of proof rules for structured rv-programs:

{\small\svsp\bd

\item{\em Basic rule} (see Fig.~\ref{ro-r1}(left)): The validity of an assertion \hass{\tau (NE)\sigma\mbox{:}
  C}{M}{\tau (EN)\sigma\mbox{:} C'} for a module $M$ is reduced to the validity of the assertion
  \hass{C}{M}{C'} in the setting of usual while programs (enriched with equalities showing that the variables
  in $\tau$ and $\sigma$ does not change by passing form $C$ to $C'$).

\item {\em Rule for horizontal composition} (see Fig.~\ref{ro-r1}(right)): If\\ \hass{\tau
  (N^kE^l)E^m\sigma\mbox{:} C}{P1}{\tau (E^lN^k)E^m\sigma\mbox{:} C1} and\\ \hass{\tau
  E^l(N^kE^m)\sigma\mbox{:} C1}{P2}{\tau E^l(E^mN^k)\sigma\mbox{:} C2}, then\\ \hass{\tau
  (N^kE^{l+m})\sigma\mbox{:} C}{P1\phcomp P2}{\tau (E^{l+m}N^k)\sigma\mbox{:} C2}.

\item {\em Rule for vertical composition:} similar

\item {\em Rule for diagonal composition:} If\\ \hass{\tau (N^kE^l)\sigma\mbox{:} C}{P1}{\tau
  (E^lN^k)\sigma\mbox{:} C1} and\\ \hass{\tau (N^kE^l)\sigma\mbox{:} C1}{P2}{\tau (E^lN^k)\sigma\mbox{:} C2},
  then\\ \hass{\tau (N^kE^l)\sigma\mbox{:} C}{P1\pdcomp P2}{\tau (E^lN^k)\sigma\mbox{:} C2}.\\ (Notice that
  $C1$ is used on two different contour lines. The above convention on variable indices ensures the rule is
  sound.)

\item {\em Rule for ``if''\mbox{:}} For $Q=if (Cond)\{P1\} else \{P2\}$, if\\ \hass{\tau
(N^kE^l)\sigma\mbox{:} C\wedge Cond}{P1}{\tau (E^lN^k)\sigma\mbox{:} C'} and\\ \hass{\tau
(N^kE^l)\sigma\mbox{:} C\wedge \neg Cond}{P2}{\tau (E^lN^k)\sigma\mbox{:} C'}, then\\ \hass{\tau
(N^kE^l)\sigma\mbox{:} C}{Q}{\tau (E^lN^k)\sigma\mbox{:} C'}.

\item {\em Rule for autonomous temporal or spatial ``while'':} For a temporal while with dummy temporal
  interfaces (i.e., the west and east interfaces have a $nil$ type), the classical while rule may be used. By
  space-time duality, a similar rule applies to a spatial while with dummy spatial interfaces.

\item {\em Rule for spatio-temporal ``while'':} If an invariant $Inv$ may be found such that\\ \hass{\tau
  (N^kE^l)\sigma\mbox{:} Inv\wedge Cond}{P}{\tau (E^lN^k)\sigma\mbox{:} C'} and $C'\ra Inv$,
  then,\\ \hass{\tau (N^kE^l)\sigma\mbox{:} Inv}{while\_st(Cond)\{P\}}{\tau (E^lN^k)\sigma\mbox{:} Inv\wedge
    \neg Cond}.\\ (See the comment on the diagonal composition to clarify the use of the assertions $C$ and
  $Inv$ on different contour lines.)

\item {\em Rule for a simple ``for'':} If {\tt i} is not changed by {\tt R} in a statement $Q=$ {\tt
for\_s(i=0;i<a;i++)\{R\}}, then the following rule applies: if\\ \hass{\tau
(E^l)^j(N^kE^l)(E^l)^{a-j-1}\sigma\mbox{:} C_j}{R}{\tau (E^l)^j(E^lN^k)(E^l)^{a-j-1}\sigma\mbox{:} C_{j+1}},\\
for all $j<a$, then\\ \hass{\tau (N^k(E^l)^{a-1})\sigma\mbox{:} C_0}{Q}{\tau ((E^l)^{a-1}N^k)\sigma\mbox{:}
C_{a-1}}.

\item {\em Rule for implication:} For a Hoare assertion \hass{\tau (N^kE^l)\sigma\mbox{:} C}{P}{\tau
(E^lN^k)\sigma\mbox{:} C'}, if $D\ra C$, $C'\ra D'$, then \hass{\tau (N^kE^l)\sigma\mbox{:} D}{P}{\tau
(E^lN^k)\sigma\mbox{:} D'}.\ed}

\bthm The inference rules are sound, i.e., if an assertion \svsp\\\centerline{\hass{\tau
    (N^kE^l)\sigma\mbox{:} C}{P}{\tau (E^lN^k)\sigma\mbox{:} C'}}\svsp\\ is proved, then all scenarios of $P$
satisfying the input condition satisfy the output condition, too.\hfill{$\Box$}\ethm

\section{A case study: The verification of a distributed termination detection protocol}\label{s-ring}

As a case study, we verify the correctness of a termination detection protocol. The activity of processes and
their interactions are all described using structured rv-programs.

Termination detection is quite a popular research topic in distributed systems. The aim is to find when a set
of distributed processes have terminated. The problem is particularly complicate as one has to combine a local
termination condition (each process has finished its current jobs) with a global condition (no messages are in
transit, as such messages may reactivate already terminated processes). There are many termination detection
protocols - we study a popular (dual-pass) ring termination detection protocol (see, e.g., \cite{dij87}).

\paragraph{Ring termination detection.}

The dual-pass ring termination detection protocol is used to detect the termination of a pool of distributed
processes logically organized as a ring.  The protocol can handle the case when processes may be reactivated
after their local termination. To this end, it uses colored (i.e., black or white) tokens. Processes are also
colored: a black color means global termination may have not occurred. The algorithm works as follows:

(1) The root process $P_0$ becomes white when it has terminated and it generates a white token that is passed
to $P_1$.

(2) The token is passed through the ring from one process $P_i$ to the next when $P_i$ has
terminated. However, the color of the token may changed. If a process $P_i$ passes a task to a process $P_j$
with $j<i$, then it becomes a black process; otherwise it is a white process. A black process will pass on a
black token, while a white process will pass on the token in its original color. After $P_i$ has passed on a
token, it becomes a white process.

(3) When $P_0$ receives a black token, it passes on a white token; if it receives a white token, all
processes have terminated.

\subsection{Implementation}
 
Suppose there are ${\tt n}$ processes, denoted {\tt 0,\dots,n-1}.  Besides the input ${\tt n}$, the program
uses the spatial variables ${\tt id : sInt,\ c : \{white,black\},\ active : sBool}$ and the temporal variables
${\tt tn, tid : tInt,\ msg : tIntSet[~]}$. Their role is described below.

\begin{figure}${}$\hfill{\includegraphics[scale=.4]{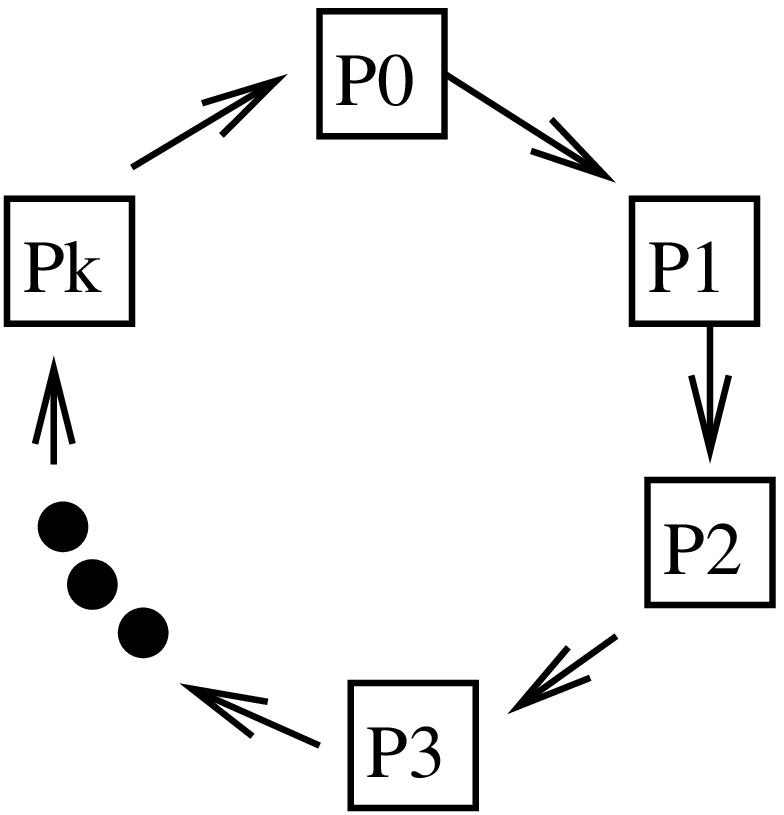}}\vspace{-4.cm}\\ 
\begin{center}\hspace*{-2cm}A \defi{run} (for termination detection program)\\
\includegraphics[scale=.4]{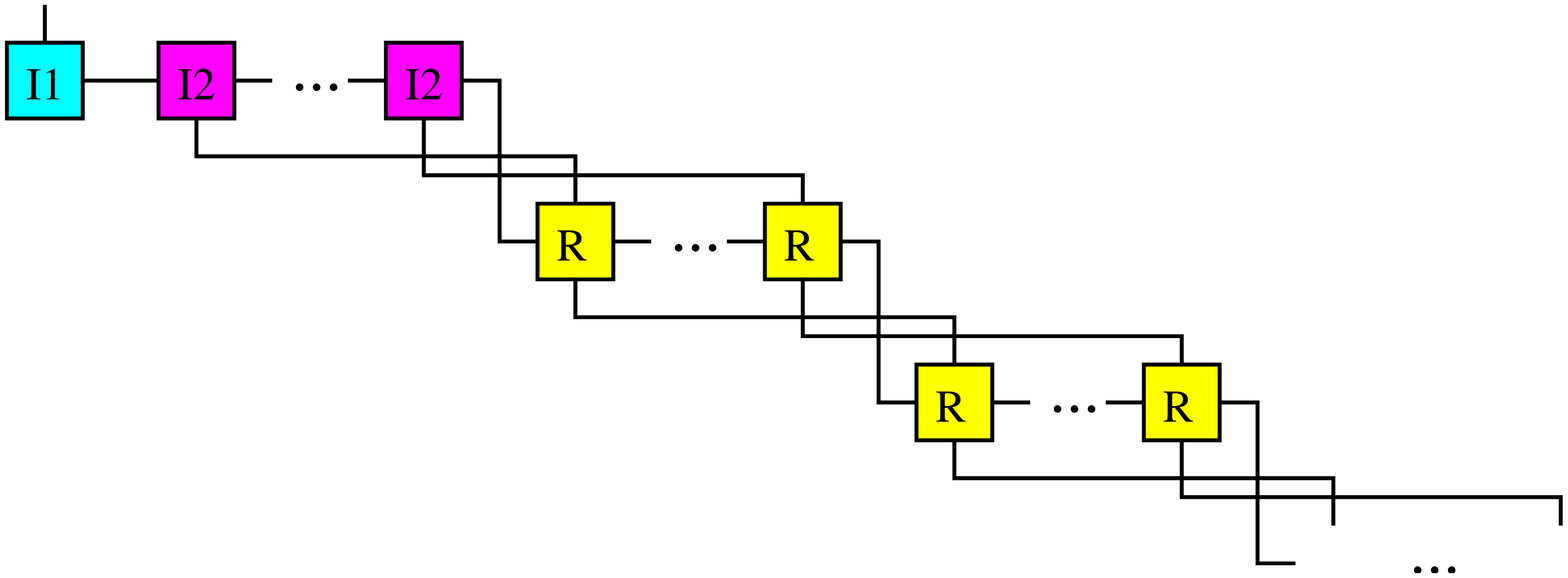}\vspace{-.4cm}\end{center}
\caption{Vertical and horizontal compositions and ``if'' statements}
\label{scen-termin}\end{figure}

Our structured rv-program ${\tt P}$ implementing the dual-pass ring termination protocol is the diagonal
composition of an initialization program ${\tt I}$ and a core program ${\tt Q}$, \snvsp$${\tt
  P=I\ \pdcomp\ Q}\snvsp$$ where \bi\item[]\bi

\item[${\tt I={}}$] {\tt\small I1\phcomp\ for\_s(tid=0;tid<tn;tid++)\{I2\}\phcomp}

\item[${\tt I1={}}$]{\tt\small module\{listen nil\}\{read n\}\{ \\
tn=n; token.col=black; token.pos=0; \\ 
\}\{speak tn,tid,msg[~],token(col,pos)\}\{write nil\}}

\item[${\tt I2={}}$]{\tt\small module\{listen tn,tid,msg[~],token(col,pos)\}\{read nil\}\{ \\ 
id=tid; c=white; active=true; msg[id]=null;\\ 
\}\{speak tn,tid,msg[~],token(col,pos);\}\{write id,c,active\}}

\item[${\tt Q={}}$]{\tt\small while\_st(!(token.col==white \&\& token.pos==0))\{\\
\hspace*{.5cm} for\_s(tid=0;tid<tn;tid++)\{R\}\}}

\item[${\tt R={}}$]{\tt\small
module\{listen tn,tid,msg[~],token(col,pos)\}\{read id,c,active\}\{ \\ 
for(j=0;j<tn;j++)\{ //take my jobs\\
\hspace*{.5cm}if(msg[j] contains id)\{\\ 
\hspace*{1.cm} msg[j]=msg[j]-\{id\};\\
\hspace*{1.cm} active=true;\};\}\\
if(active)\{ //execute code, send jobs, update color\\
\hspace*{.5cm} delay(random\_time);  \\
\hspace*{.5cm} r=random(tn-1); \\ 
\hspace*{.5cm} for(i=0;i<r;i++)\{ k=random(tn-1); \\
\hspace*{1.cm} if(k!=id)\{msg[id]=msg[id]$\cup$\{k\}\};\\ 
\hspace*{1.cm} if(k<id)\{c=black\};\}\\ 
\hspace*{.5cm} active=random(true,false);\}\\ 
if(!active \&\& token.pos==id)\{ //termination\\
\hspace*{.5cm} if(id==0)token.col=white;\\
\hspace*{.5cm} if(id!=0 \&\& c==black)\{\\
\hspace*{1.cm} token.col=black;c=white\};\\
\hspace*{.5cm} token.pos=token.pos+1[mod tn];\}\\
\}\{speak tn,tid,msg[~],token(col,pos);\}\{write id,c,active\}}
\ei\ei

Notice that, except for the operations on sets (for which AGAPIA programs have to be provided), the code
represent a valid AGAPIA v0.1 program.

\paragraph{Comments.}

The spatial variables ${\tt id,c,active}$ represent the process identity, its color, and its active/passive
status.  The temporal variables used in this program are: (i) ${\tt tn,tid}$ - temporal versions of ${\tt
n,id}$; (ii) ${\tt msg[~]}$ - an array of sets, where ${\tt msg[k]}$ contains the ${\tt id}$ of the
destination processes for the pending messages sent by process ${\tt k}$; (iii) ${\tt token.col}$ - an element
of ${\tt \{white,black\}}$ representing the color of the token; and (iv) ${\tt token.pos}$ - the number of the
process that has the token.

The program starts with the initialization of the network (program ${\tt I}$) by activating all the processes
(and setting the fields ${\tt id,c,active}$).  Initially, ${\tt msg[i]=\emptyset}$, for all ${\tt 0\leq i <
n}$, because no jobs were sent and the default color/position of the token is black/0.

After the initialization part and until the first process receives a white token back, each process executes
its code. If one process has the token and terminates, it passes the token to the next process (only the first
process has the right to change the color of the token into white once it terminates).

When a process executes the code ${\tt R}$, whether active or passive, it checks if new jobs were assigned to
it; if the answer is positive, it collects its jobs from the jobs lists and stays/becomes active. When it is
active, it executes some code, sends new jobs to other processes, and randomly goes to an active or passive
state. If it has the token, it keeps it until it reaches termination and afterward it passes it. A white
process will pass the token with the same color as it was received and a black process will pass a black token
(after passing the token, the process becomes white).\vspace*{-0.1cm}

\subsection{Verification}

The program ${\tt P}$ is the diagonal composition of the initialization block and the repeated diagonal
compositions given by the ${\tt while\_st}$ statement. A typical run is presented in
Fig.~\ref{scen-termin}. In each case, the temporal/spatial output of a block becomes the temporal/spatial
input of the next block.

For ${\tt I}$, the input is a spatial variable ${\tt n}$. The output satisfies the condition:\bi\item[]
$\forall k\in[0,n)\co (id,c,active)[k]=(k,white,true)\\ \wedge tn=n\wedge token=(black,0) \wedge
\forall k\in[0,n)\co msg[k]=\emptyset$.\ei Notice that the spatial interface is expanded on ${\tt n}$
processes ${\tt 0,1,\dots,n-1}$. The notation ${\tt (id,c,active)[k]}$ refers to the the values of variables
${\tt (id,c,active)}$ in process ${\tt k}$.

\paragraph{The invariant $Inv$.}

For ${\tt Q}$ we need to find appropriate invariant properties. We define the following properties and prove
they are satisfied by the program {\tt for\_s(tid=0;tid<tn;tid++)\{R\}}:\snvsp
\bi\item[]\bi\item[P1:]$token=(white,i)\ra \\
\hspace*{.5cm}[(\forall r\in[0,i-1]\co active[r]=false\wedge msg[r]=\emptyset)\\
\hspace*{.5cm}\vee(\exists k>i-1\co c[k]=black)]$\\ where the value $i-1$ is interpreted as $tn-1$ for
$i=0$.\snvsp\ei\ei In words, if the token is white and reached process $i$, then all processes with smaller {\tt
id} terminate and have no pending messages sent\foo{The pending message lists are the lists of messages that
have been inserted in and not removed from the message lists during a complete passing through the
ring. Formally, they are $msg[r]$'s at the start of the {\tt for\_s} statement.} or a process with a larger
{\tt id} is black.  \snvsp 
\bi\item[]\bi\item[P2]$token.col=white\ra(\forall k\in [0,n)\co msg[k]\not=\emptyset\ra c[k]=black)$\snvsp\ei\ei 
In words, if a process has a job inserted in its pending message list, then its color is black.\svsp

We want to prove $Inv={\tt P1\wedge P2}$ is really an invariant, i.e., the same assertion $Inv$, translated to
the output values of the variables, holds at the end of the {\tt for\_s} statement. Formally,
\snvsp$$\{|Inv|\} \mbox{\tt for\_s(tid=0;tid<tn;tid++)\{R\}} \{|Inv|\}.\snvsp$$ Notice that due to the fact
that the token is black, $Inv$ holds at the beginning of the spatio-temporal while.

\paragraph{Proof of the invariance of $Inv$.}

To simplify the presentation, we directly prove the invariance of $Inv$ for \mbox{\tt
for\_s(tid=0;tid<tn;tid++)\{R\}}. A fully formal proof, including an appropriate new invariant for {\tt R}
itself, in included in Appendix A.

Suppose $Inv$ holds at the start of the ${\tt for\_s}$ statement. We want to prove that the property
$Inv'={\tt P1'\wedge P2'}$, where $Inv'$ is $Inv$ translated to the output values of the variables\foo{
We use the standard ``prim'' notation, i.e., if $x$ is a variable, then $x'$ refers to the value of the
variable $x$ at the end of the program. Here, the convention applies also to $i$ (which is not directly a
variable of the program, but it actually denotes $token.pos$).}, holds at the end of the ${\tt for\_s}$
statement.

First, we prove ${\tt P1'}$, where\snvsp
\bi\item[]\bi\item[${\tt P1':}$] $token'=(white,i')\ra \\
\hspace*{.2cm}[(\forall r\in[0,i'-1]\co active'[r]=false\wedge msg'[r]=\emptyset)\\
\hspace*{.2cm}\vee(\exists k>i'-1\co c'[k]=black)]$\snvsp\ei\ei

Suppose $token'.col=white$; then $token.col=white$, too.  Notice that $\forall r\in[i,i'-1]\co
active'[r]=false\wedge msg'[r]=\emptyset$ holds because: (i) the token could not reach the process $i'$ unless
processes $i,\dots,i'-1$ hadn't terminated and (2) $token'.col$ hadn't been white unless
$msg'[i],\dots,msg'[i'-1]$ are all empty.

As ${\tt P1}$ holds and $token=(white,i)$, either (i) or (ii) below applies, where:

(i) $\forall r\in[0,i-1]\co active[r]=false\wedge msg[r]=\emptyset$: In this case:\\ (a) If all
processes $0,\dots,i-1$ stay passive, then by the above observation this situation is extended to $\forall
r\in[0,i'-1]\co active'[r]=false\wedge msg'[r]=\emptyset$ and we are done.\\ (b) If one process
$0,\dots,i-1$ becomes active, it may be reactivated only by a message from a process $k$ with $k>i-1$ (indeed,
$msg[0],\dots,msg[i-1]$ are all empty). Then, by ${\tt P2}$, $c[k]=black$. Moreover $k>i'-1$ (otherwise
$token'.col$ hadn't been white), hence $c'[k]=black$ and the second part is true.

(ii) $\exists k>i-1\co c[k]=black$: In this case, $k>i'-1$ (otherwise $token'.col$ hadn't been white)
and $c'[k]=black$, hence the implication holds.\svsp

Next, we prove ${\tt P2'}$, where\snvsp\bi\item[]\bi
\item[${\tt P2':}$] $token'.col=white\ra(\forall k\in [0,n):msg'[k]\not=\emptyset\ra c'[k]=black)$\snvsp\ei\ei

Notice that after the execution of ${\tt R}$ by the process $k$, $msg'[k]$ consists in the processes that were
contacted by $k$.  The execution of ${\tt R}$ for $tid=k$ is followed by the execution of ${\tt R}$ for
$k<tid<tn$.  All these executions of ${\tt R}$ that follows, will discard all the messages sent to processes
greater than $k$ from $msg'[k]$ and consequently, by the end of the ${\tt for\_s}$, $msg'[k]\subseteq [0,k)$.

Hence, if $msg'[k]\not=\emptyset$, then the process $k$ had sent a message to a process $p$ with $p<k$ and the
color of the process became black. Moreover, if $token'.col=white$, then the color of the process stayed black
until the end of the ${\tt for\_s}$ instruction, which implies $c'[k]=black$.

\paragraph{The final step.}

Applying the rule for the spatio-temporal while \snvsp\bi
\item[]$\{|Inv|\}\mbox{\tt while\_st(!(token=(white,0)))\{Q'\}}\{|Inv \wedge (token=(white,0))|\}$\snvsp\ei
where ${\tt Q'= \mbox{\tt for\_(tid=0;tid<tn;tid++)\{R\}}}$, it follows that \snvsp$$\forall i\in[0,tn-1]\co
active[i]=false\wedge msg[i]=\emptyset\snvsp$$ hence all process have terminated and there are no pending
jobs/messages in the communication lists.

\bthm The program for the dual-pass ring termination detection protocol is
correct.\hfill{$\Box$}\ethm

It would be interesting to compare our proof with proofs of the protocol using process algebra or other formal
verification methods, if such proofs are available.

\section{Related and future works}\label{s7}

This is a brief section on related works, with emphasis on two-dimensional patterns and spatio-temporal
logics. Our grids (scenarios without data around) are closely related to two-dimensional (or picture) languages
\cite{gi-re97,lmn98}\foo{Interesting extensions (logics and calculi) for modeling continuous two-dimensional
  shapes are described in \cite{car08,sch04}.} - actually, finite interactive systems \cite{ste02} (on which
rv-systems are based), are equivalent to tile systems or to existential monadic second order logic
\cite{grst96}. Regarding scenarios, a worthwhile approach may be to use results on two-dimensional languages
in combination with model-checking to (lightly) verify rv-programs.

Space and time are fundamental entities, so no surprise to find many proposals on developing space-time
logics. Compared with \cite{ca-go00}, we use linear not branching space and time. Tile logic
\cite{bru99,ga-mo99} use similar two-dimensional patterns, but with emphasis on rewriting and declarative
computation models. Other interesting space-time proposals on verifying mobile or open systems are presented
in \cite{mwz03,va-ag01}.

\newpage
\section*{Appendix A: On the termination detection protocol}

\paragraph{A fully formal proof of $Inv$ invariance.}

Here we give a fully formal proof (in $STHlog_0$ for $$\{|Inv|\} \mbox{\tt for\_s(tid=0;tid<tn;tid++)\{R\}}
\{|Inv|\}$$ For the beginning, we identify a few more detailed assertions, which essentially depend on ${\tt
  tid}$ and describe the effect of the computation within ${\tt R}$: \bi
\item[]${\tt Q1:}$ (case $token.col=white
\wedge tid<token.pos$)\\ $token'=token\wedge\forall k\not=tid\co msg'[k]=msg[k]-\{tid\}\wedge \\
\hspace*{.5cm}[\hsp (active[tid]=false \wedge {}\not\exists k, tid\in msg[k]\\
\hspace*{1.5cm}\ra active'[tid]=false\wedge msg'[tid]=\emptyset)\\
\hspace*{.5cm}\vee (active[tid]=true \vee\exists k, tid\in msg[k]\\
\hspace*{1.5cm}\ra msg'[tid]\subseteq[0,tid)\cup[tid+1,n)\wedge \\
\hspace*{2.1cm}msg'[tid]\cap [0,tid)\not=\emptyset\ra c'[tid]=black)]$
\svsp\\${\tt Q2:}$  (case $token.col=white \wedge tid=token.pos$)\\
$\forall k\not=tid\co msg'[k]=msg[k]-\{tid\}\wedge\\
\hspace*{.5cm}[\hsp(active'[tid]=true \\
\hspace*{1.5cm} \ra token'=token\wedge \\
\hspace*{2.1cm} msg'[tid]\cap [0,tid)\not=\emptyset\ra c'[tid]=black)\\
\hspace*{.5cm}\vee (active'[tid]=false\\ 
\hspace*{1.5cm} \ra token'.pos=token.pos+1\wedge \\ 
\hspace*{2.1cm}token'.col=white \ra msg'[tid]\cap [0,tid)=\emptyset)]$
\svsp\\${\tt Q3:}$ (case $token.col=white \wedge tid>token.pos$) - same as {\tt Q1}.\ei

To prove $Inv$, we use a more detailed version $Inv2$ satisfying the following properties: (i) If $Inv2$ holds
``up-to'' to an $tid$ and module ${\tt R}$ is applied, then $Inv2$ holds up to $tid+1$; (ii) $Inv$ follows
from the fact that $Inv2$ holds for the last value of $tid$. Formally, we have to prove \bi\item[]\bi
\item[]$\{|Inv2|\}{\tt\ R\ }\{|Inv2'|\}$;\\ $Inv2\wedge (tid=n)\ra Inv$\ei\ei The basic step is illustrated in the
next figure (Fig.~\ref{f-hoare-step}). 

\begin{figure}[bh]\begin{center}\includegraphics[scale=.35]{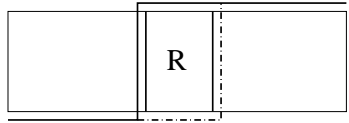}\end{center}
\caption{The basic step for the application of Hoare method: a classical triple surrounding $R$ extended with
  an empty contour\vspace{-.3cm}}
\label{f-hoare-step}\end{figure}

The new invariant $Inv2$ is ${\tt P1d}\wedge {\tt P2d}$, where {\tt P1d,P2d} are the following slightly more
detailed variations of the previous properties {\tt P1,P2}:\bi\item[]\bi
\item[${\tt P1d:}$]$token=(white,i)\ra \\
\hspace*{.5cm}[(\forall r\in[0,i-1]\co active[r]=false\wedge msg[r]\subseteq[max(tid,i),n))\\
\hspace*{.5cm}\vee(\exists k>i-1\co c[k]=black)]$\\ where, as before, the value $i-1$ is interpreted as $tn-1$
for $i=0$.
\item[${\tt P2d:}$] $token.col=white\ra\\ \hspace*{.5cm}\forall k\in [0,tid)\co $\parbox[t]{8cm}{
$msg[k]\subseteq [0,k)\cup[tid,n) \wedge \\ msg[k]\cap[0,k)\not=\emptyset\ra c[k]=black$}\ei\ei

\paragraph{Proof of {\tt P1d}.} 

For $\{|{\tt P1d}|\}{\tt\ R\ }\{|{\tt P1d}|\}$, we have to prove that {\tt P1d} and {\tt Q1-3} implies
\bi\item[]\bi\item[${\tt P1d':}$]$token'=(white,i')\ra \\
\hspace*{.5cm}[(\forall r\in[0,i'-1]\co active'[r]=false\wedge msg'[r]\subseteq[max(tid',i'),n))\\
\hspace*{.5cm}\vee(\exists k>i'-1\co c'[k]=black)]$\ei\ei 

Suppose $token'=(white,i')$ and $i'=i$, hence: $tid\not=i$ or ($tid=i$ and process $tid$ doesn't
terminate). By {\tt P1d}, either (i) or (ii) holds, where\bi
\item[(i)] $\exists k>i-1\co c[k]=black$: The property is preserved, i.e., $c'[k]=black$. (Indeed, a black
process may become white only if it terminates, which is not the case as $i'=i$.)
\item[(ii)] $\forall r\in[0,i-1]\co active[r]=false\wedge msg[r]\subseteq[max(tid,i),n)$ and $\neg(\exists
k>i-1\co c[k]=black)$: For the ``active'' part: \bi
\item If $tid\geq i$ the property is outside of the action of {\tt R}, hence still true. 
\item If $tid<i$, the process $tid$ cannot be activated by a processes $r$ with $r<i$ (there are no messages
for $tid$ there). On the other hand, if a process $r$, with $r\geq i$, activates process $tid$, then by $P2d$
its color $c[r]$ is black and this contradicts this care premises. \ei For the second part, notice that a
process $r$ with $r\not=tid$ has $msg'[r]=msg[r]-\{tid\}$, while the process $tid$ with $tid<i$ is inactive,
hence $msg'[tid]=\emptyset$.  \ei

Suppose $token'=(white,i')$ and $i'=i+1$, hence: $tid=i$, the process $tid$ terminates, and $tid$ was a white
process before termination. Again by {\tt P1d}, either (i) or (ii) holds, where\bi
\item[(i)] $\exists k>i-1\co c[k]=black$: As $token'=(white,i')$, actually $k>i$, hence the property is
preserved.
\item[(ii)] $\forall r\in[0,i-1]\co active[r]=false\wedge msg[r]\subseteq[max(tid,i),n)$ and $\neg(\exists
k>i-1\co c[k]=black)$: For $r<i$ the proof is as before. For $r=i$, by the previous observations,
$active'[i]=false\ \wedge\ msg'[i]\subseteq[i+1,n)$, hence the property is preserved\foo{Notice that $tid=i$,
hence $i+1=i'=tid'=max(tid',i')$.}.  \ei
 
Finally, notice that at the end of the ${\tt for\_s}$ statement $tid=n$; moreover, clearly ${\tt P1d}\wedge
(tid=n)\ra {\tt P1}$.

${}$\vspace{-.8cm}\paragraph{Proof of {\tt P2d}.}

For $\{|{\tt P2d}|\}{\tt\ R\ }\{|{\tt P2d}|\}$, we have to prove \bi\item[]\bi
\item[${\tt P2d':}$] $token'.col=white\ra\\ \hspace*{.5cm}\forall k\in [0,tid')\co $\parbox[t]{8cm}{
$msg'[k]\subseteq [0,k)\cup[tid',n) \wedge \\ msg'[k]\cap[0,k)\not=\emptyset\ra c'[k]=black$}\ei\ei 
This directly follows from {\tt P2d} and {\tt Q1-3}\foo{In the last implication, if
$msg'[k]\cap[0,k)\not=\emptyset$ the process becomes black by the first part of the code of ${\tt R}$. Its
color may be changed to white by the last part of the code of ${\tt R}$ only if the process has the token and
terminates, but then the token will be black and this contradicts the premise $token'.col=white$.}. Finally,
after ${\tt for\_s}$ statement $tid=n$ and ${\tt P2d}\wedge (tid=n)\ra {\tt P2}$.

\bthm The program for the dual-pass ring termination detection protocol is correct. Moreover, there is a fully
formal proof of its correctness using the $STHlog_0$ inference rule defined in
Sec.~\ref{s-hoare}.\hfill{$\Box$}\ethm

\end{document}